\documentclass[aps,prl,twocolumn,showpacs,floatfix]{revtex4}

\usepackage{graphicx}
\usepackage{dcolumn}
\usepackage{bm}

\begin{document}

\title{Fine Structure of the QCD String Spectrum }

\author{K.~Jimmy Juge}
\affiliation{Institute for Theoretical Physics, University of Bern,
   Sidlerstrasse 5, CH-3012 Bern, Switzerland}
   
\author{Julius Kuti}
\affiliation{Department~of Physics, University of California at San Diego,
    La Jolla, California 92093-0319} 

\author{Colin Morningstar}
\affiliation{Department~of Physics, Carnegie Mellon University,
    Pittsburgh, PA 15213, USA}

\date{July 10, 2002}

\begin{abstract}
Using advanced lattice methods in Quantum Chromodynamics,
three distinct scales are established in the
excitation spectrum of the gluon field around a static 
quark-antiquark pair as the color source separation $R$ is varied. 
On the shortest length scale, the excitations are consistent 
with states created by local gluon field operators
arising from a multipole operator product expansion.
An intermediate crossover region below 2 fm is identified 
with a dramatic rearrangement of the level orderings.
On the largest length scale of 2-3 fm, the spectrum agrees
with that expected for string-like excitations.
The energies nearly reproduce asymptotic $\pi/R$ string gaps,
but exhibit a fine structure, providing important clues for developing
an effective bosonic string description. 
\end{abstract}


\maketitle

There exists great interest and considerable effort to explain quark 
confinement in Quantum Chromodynamics
from the string theory viewpoint. The ideas of
't~Hooft, Polyakov, Witten, and others, and 
recent glueball spectrum and string tension
calculations in AdS theories are illustrative examples.
In a somewhat complementary approach, the search for a microscopic
mechanism to explain quark confinement in the QCD vacuum continues with
vigorous effort. 
Our view is that the relevant properties of the underlying 
effective QCD string theory,  
whether it emerges from strictly string theoretic ideas 
or from the microscopic
theory of the vacuum, are coded in the excitation spectrum of the
confining flux. To establish the main features of this spectrum is
the objective of this work. 

We report here, for the first time, our comprehensive determination 
of the rich low-lying energy spectrum of the excited gluon field
between a static quark--antiquark $(q\bar q)$ pair in the fundamental 
color representation of $SU(3)_c$.
Only a few of these states have been previously studied and
only for $q\bar q$ separations $R$ up to 1 fm~\cite{ford,perantonis,bali}.
Due to our use of recent advances in lattice gauge theory technology,
which include anisotropic lattices, 
an improved lattice gauge action, and large sets of creation
operators, we have been
able to determine the excitation spectrum for 
a large range of $R$ values from 0.1 to 3 fm, and for a greatly extended
set of excited states never studied before.
To investigate the onset
of string-like behavior when the flux is much longer
than its intrinsic width became a realistic goal~\cite{JKM2}. 
The effects of light $q\bar q$ pairs in the vacuum are
not included in this study, with focus on the confining properties
of the gluon field.
Earlier results from less extensive simulations were reported in  
Ref.~\cite{JKM1}.

Three exact quantum numbers which are based on the symmetries of the problem
determine the classification scheme of the gluon excitation spectrum
in the presence of a static $q\bar q$ pair.
We adopt the standard notation from the physics of diatomic molecules
and use $\Lambda$ to denote the magnitude of the eigenvalue of the projection
${\bf J}_g\!\cdot\hat{\bf R}$ of the total angular momentum ${\bf J}_g$
of the gluon field onto the molecular axis with unit vector
$\hat{\bf R}$. The capital Greek
letters $\Sigma, \Pi, \Delta, \Phi, \dots$ are used to indicate states
with $\Lambda=0,1,2,3,\dots$, respectively.  The combined operations of
charge conjugation and spatial inversion about the midpoint between the
quark and the antiquark is also a symmetry and its eigenvalue is denoted by
$\eta_{CP}$.  States with $\eta_{CP}=1 (-1)$ are denoted
by the subscripts $g$ ($u$).  There is an additional label for the
$\Sigma$ states; $\Sigma$ states which
are even (odd) under a reflection in a plane containing the molecular
axis are denoted by a superscript $+$ $(-)$.  Hence, the low-lying
levels are labeled $\Sigma_g^+$, $\Sigma_g^-$, $\Sigma_u^+$, $\Sigma_u^-$,
$\Pi_g$, $\Pi_u$, $\Delta_g$, $\Delta_u$, and so on.  For convenience,
we use $\Gamma$ to denote these labels in general.

The gluon excitation energies $E_\Gamma(R)$ were extracted from 
Monte Carlo estimates of generalized large Wilson loops using
extended sets of 15 to 25 operators, projected onto the various
symmetry sectors, on three-dimensional
time slices of the lattice.
Monte Carlo estimates of the $W_\Gamma^{ij}(R,t)$ Wilson
correlation matrices were obtained
in a series of simulations performed on 
our dedicated UP2000 Alpha cluster using an improved gauge-field
action~\cite{MP1}.  The couplings $\beta$, input aspect ratios $\xi$,
and lattice sizes for each simulation will be detailed in 
forthcoming publications but are available now
upon request.  
The use of anisotropic lattices, with
the temporal lattice spacing $a_t$ much smaller than the spatial
spacing $a_s$, was crucial for resolving the gluon
excitation spectrum, particularly
for large $R$. The renormalized anisotropy was determined using the torelon
dispersion relation.
To hasten the onset of asymptotic behavior in effective mass plots, 
iteratively-smeared spatial links were used in the generalized 
Wilson loops whose temporal segments  were constructed from
thermally-averaged links to reduce statistical noise.

\begin{figure}
\includegraphics[width=3.45in,bb=50 70 554 1130]{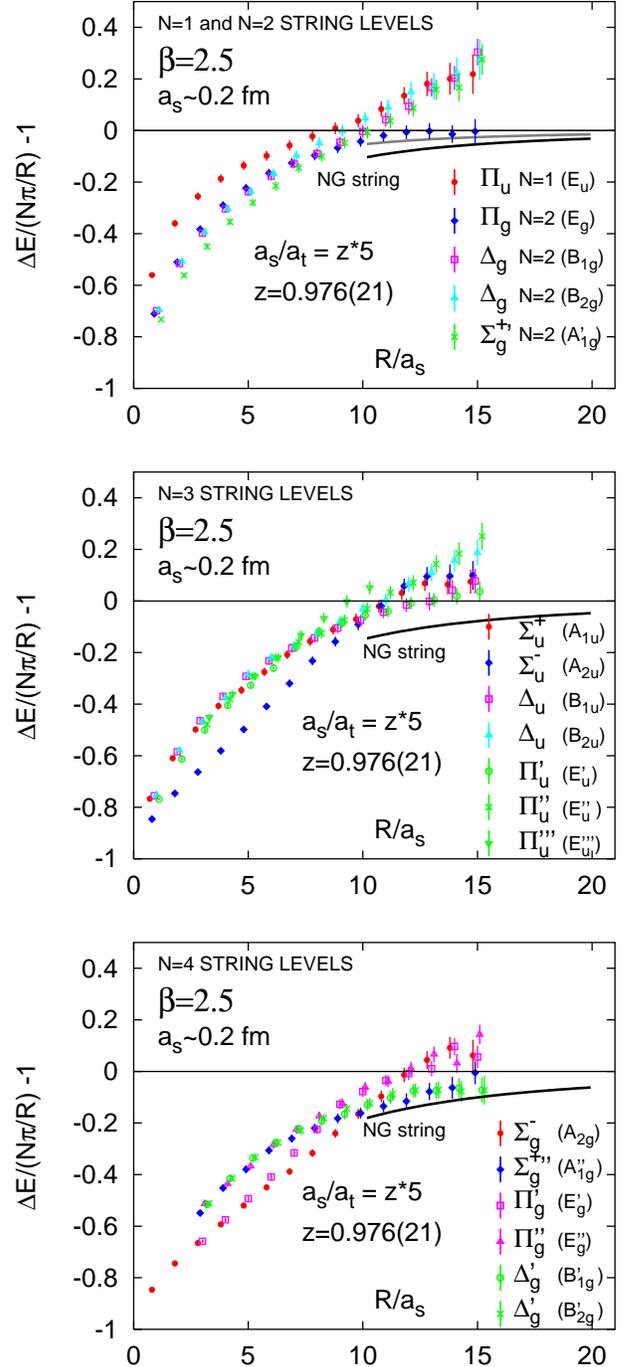}
\caption{\label{fig:fig1}
Energy gaps $\Delta E$ above $\Sigma^+_g$  are shown
in string units for quantum numbers
in continuum and lattice notation. 
The Nambu-Goto string is discussed in the text.
}
\end{figure}

Restricted to the $R=0.2-3$ fm range of a selected simulation,
energy gaps $\Delta E$ above the ground state are compared to
asymptotic string gaps for 18 excited states 
in Fig.~\ref{fig:fig1}.
The quantity $\Delta E/(N\pi/R) - 1$ is plotted to show percentage deviations   
from the asymptotic string levels for string quantum number $N=1,2,3,4$.
The energy gaps, far below the null lines of 
the plots and strongly split at fixed $N$,
differ from the simple string gaps for $R<2$~fm and a fine structure remains
visible beyond $R$ = 2 fm.
The three relevant scales of the spectrum are discussed below.

The errors shown in Fig.~1 are statistical only and were determined
using standard bootstrap techniques.  Autocorrelations were explicitly
calculated to ensure that the statistical errors were not underestimated.
Dependence of the error estimates on the choice of temporal fit ranges
was minimized by using fits to sums of two exponentials.  The removal of
contamination in our results from excited-states was checked using
a variety of methods, including Bayesian constrained fits to large
sums of exponentials, variational re-diagonalizations, and correlation
matrix fits to sums of coupled exponentials.
The insignificance of systematic effects from finite volume was
verified, and simulations
carried out for several other lattice spacings indicate that finite
spacing errors in the results shown in Fig.~1 are expected to be small.

\noindent{\bf Short distance spectrum.} 
For $R \ll 1$~fm, the observed
level ordering is consistent with the short-distance physics
of gluon field excitations which are trapped around
a dominantly color octet static $q\bar q$ pair.  
Since gluon field dynamics will only
depend softly on ${\bf R}$,
a new approximate symmetry is expected in the
$R\sqrt\sigma\ll 1$ limit (the string tension $\sigma$ sets the scale
of transverse dimension across the flux).
This symmetry was first noted 
within the context of the static bag picture of gluon
excitations~\cite{Hasenfratz} 
and used more extensively in Refs.~\cite{JKM2,JKM4,foster,soto}.
In particular, it was shown in Ref.~\cite{soto} how the
short-distance operator product expansion of gluon excitations around 
static color sources,
in lowest order of the multipole expansion of the gluon field,
can be applied to the spectrum yielding similar results to the bag
picture~\cite{JKM4} without phenomenological model assumptions.
\begin{table}[h]
\caption{\label{tab:table1} Operators to create excited gluon states
for small $q\bar q$ separation $R$ are listed. 
${\bf E}$ and ${\bf B}$ denote the electric
and magnetic operators, respectively. The
covariant derivative ${\bf D}$ is defined 
in the adjoint representation~\cite{soto}.}
\begin{ruledtabular}
\begin{tabular}{l l l}
gluon state & $J$ & operator \\
\hline
$\Sigma_g^{+\, \prime}$ & 1 & ${\bf R}\cdot{\bf E}, \quad
             {\bf R}\cdot({\bf D}\times {\bf B})$   \\
$\Pi_g$ & 1 & ${\bf R}\times{\bf E}, \quad
             {\bf R}\times({\bf D}\times {\bf B}) $ \\ \hline
$\Sigma_u^-$ & 1 & ${\bf R}\cdot{\bf B} , \quad
             {\bf R}\cdot({\bf D}\times {\bf E})$ \\
$\Pi_u$ & 1 & ${\bf R}\times{\bf B}, \quad
             {\bf R}\times({\bf D}\times {\bf E})$\\ \hline     
$\Sigma_g^-$ & 2 & $({\bf R}\cdot {\bf D})({\bf R}\cdot {\bf B})$\\   
$\Pi_g^{\prime}$  & 2 & ${\bf R}\times(({\bf R}\cdot{\bf D}) {\bf B} 
              +{\bf D}({\bf R}\cdot{\bf B}))$ \\
$\Delta_g$ & 2 & $({\bf R}\times {\bf D})^i({\bf R}\times{\bf B})^j 
             +({\bf R}\times{\bf D})^j({\bf R}\times{\bf B})^i$\\ \hline
$\Sigma_u^{+}$ & 2 & $({\bf R}\cdot {\bf D})({\bf R}\cdot {\bf E})$\\
$\Pi_u^{\prime}$ & 2 &  ${\bf R}\times(({\bf R}\cdot{\bf D}) {\bf E} 
              +{\bf D}({\bf R}\cdot{\bf E})) $\\
$\Delta_u$  & 2 & $({\bf R}\times {\bf D})^i({\bf R}\times{\bf E})^j 
             +({\bf R}\times{\bf D})^j({\bf R}\times{\bf E})^i$\\
\end{tabular}
\end{ruledtabular}
\end{table}

\begin{figure}[h]
\includegraphics[width=3.5in,bb=50 70 554 770]{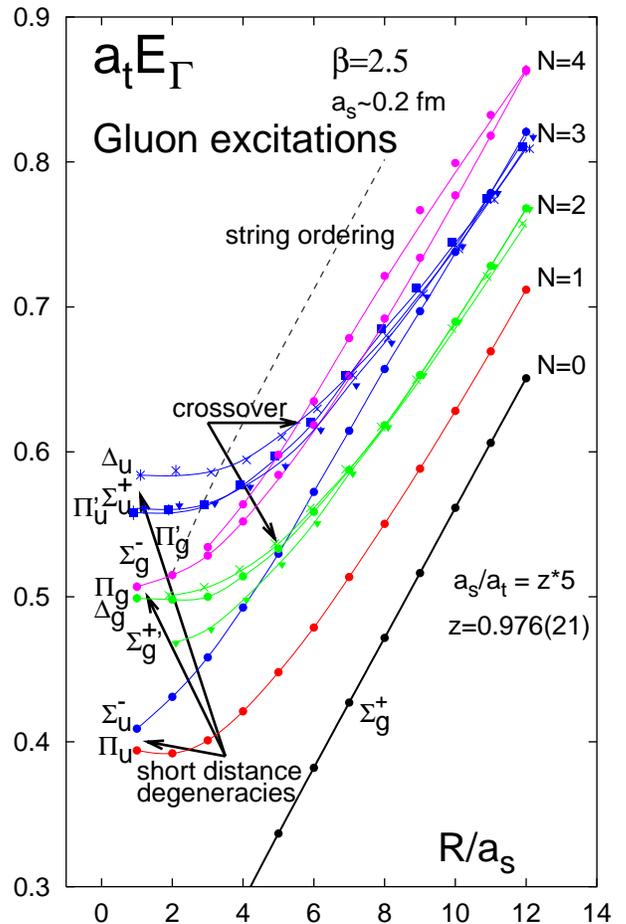}
\caption{\label{fig:fig2} Short-distance degeneracies and crossover
in the spectrum. The solid curves are only shown for visualization.
The dashed line marks a lower bound for the onset of mixing effects
with glueball states which requires careful interpretation.
}
\end{figure}

Gluon excitations will transform 
according to representations of $O(3)$ in the short-distance limit
so that $\Lambda$ in the set $\Gamma$
will be replaced by the gluon field angular momentum $J$.
A representation with fixed $J$ contains degenerate
states  with different $\Lambda$ values.
For illustration, gluon field operators of $O(3)$ multiplets
are listed in Table~\ref{tab:table1} 
for ten states
with their short distance angular momentum quantum number $J$
for each multiplet. There are two triplets 
$(\Sigma^{+'}_g,\Pi_g)$, $(\Sigma^-_u,\Pi_u)$, and two $J=2$
multiplets $(\Sigma^-_g,\Pi^{'}_g,\Delta_g)$, 
$(\Sigma^+_u,\Pi^{'}_u, \Delta_u)$ in Table~\ref{tab:table1}. 
$\Pi$ and $\Delta$ states occur with two opposite sign chiralities $\Lambda$
within each $O(3)$ multiplet.
Within the degenerate multiplets, one finds different 
string quantum numbers $N$ mixed together.
Fig.~\ref{fig:fig2} illustrates the remarkable 
working of the predicted short--distance degeneracies.  Only the states
$\Delta_u$ and $\Sigma^{+'}_g$
show considerable soft breaking of the approximate symmetry
at the shortest $R$ values.

\noindent {\bf Crossover region.}
For $0.5$ fm $ < R < 2$~fm,
a dramatic crossover of the energy levels toward
a string-like spectrum as $R$ increases is observed.
For example, the states $ \Sigma^-_u$ with $N=3$ and
$ \Sigma^-_g $ with $N=4$ break violently away from their respective 
short-distance $O(3)$ degeneracies to approach the ordering
expected from bosonic string theory near $R \sim 2$~fm.
\begin{figure}
\includegraphics[width=3.5in,bb=50 120 554 884]{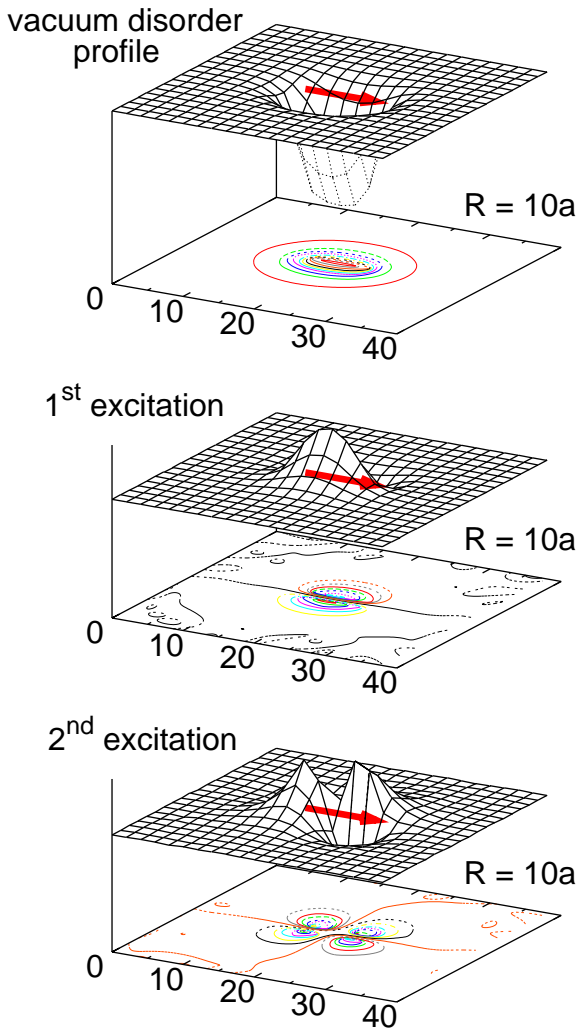}
\caption{\label{fig:fig3} 
The disorder profile of the vacuum and two
excitations are shown with arrows indicating 
the $q\bar q$ axis.}
\end{figure}

An interesting feature of the crossover region is 
the successful parametrization of the $\Sigma_{\rm g}^+$ 
ground state energy by the empirical function
$E_0(R) = a + \sigma R - c\frac{\pi}{12R}$, 
with the fitted constant $c$ 
close to unity, once $R$ exceeds $0.5$~fm.  
The Casimir energy of a thin flux line was calculated 
in Refs.~\cite{Luescher1,Luescher2}, yielding $c=1$, and this
approximate agreement
is often interpreted as evidence for string formation.
While the spectrum, including the qualitative ordering
of the energy levels, differs from the naive bosonic string gaps 
for $R < 1$~fm, a high precision 
calculation shows the rapid approach of $c_{\rm eff}(R)$  
to the asymptotic Casimir value in the same $R$ range~\cite{Luescher3}.
Although there is no inconsistency between the two 
different findings, a deeper understanding of this puzzling situation
is warranted.  

We will return to this issue in a high precision study of the 3-dimensional
$Z(2)$ gauge model in a future publication~\cite{JKM5}.  This
accurate study of $c_{\rm eff}(R)$ and the excitation spectrum of the
$Z(2)$ flux line for a wide range of $R$ values between 0.3~fm and 10~fm
will clearly demonstrate the early onset of
$c\approx 1$ without a well-developed string spectrum.  For now,
Fig.~\ref{fig:fig3} shows the lowest excitations in $Z(2)$ for $R=0.7$~fm,
revealing a bag-like disorder profile surrounding the static $q\bar q$
pair in the vacuum~\cite{JKM5}. The two lowest 
energy levels are substantially dislocated
from exact $\pi/R$ string gaps and all other excitations form
a continuous spectrum above the glueball threshold.
Since the submission of this work, a new study of
$Z(2)$ at finite temperature has appeared~\cite{caselle}, reporting 
very early onset of string behavior in support of Ref.~\cite{Luescher3}.

\noindent{\bf String limit}.
For $R > 2$~fm, the energy levels exhibit, without exception,
the ordering and approximate degeneracies of string-like excitations.
The levels nearly reproduce the asymptotic $\pi/R$ gaps, but
an intriguing fine structure remains.

It has been anticipated that the interactions of massless excitations 
on long flux lines are described by a local derivative expansion 
of a massless vector field $\bm{\xi}$ with two transverse
components in four--dimensional space-time~\cite{Luescher1, Luescher2}. 
Symmetries of the effective QCD string
Lagrangian require a derivative expansion of the form 
\begin{equation}
 {\mathcal L}_{\rm eff} = a\partial_\mu\bm{\xi}\cdot
\partial_\mu\bm{\xi} + 
b(\partial_\mu\bm{\xi}\cdot\partial_\mu\bm{\xi})^2 +
c(\partial_\mu\bm{\xi}\cdot\partial_\nu\bm{\xi})
(\partial_\mu\bm{\xi}\cdot\partial_\nu\bm{\xi}) + ..., 
\end{equation}
where the dots represent further terms with four or more derivatives
in world sheet coordinates.
The coefficient $a$ has the dimension of a mass
squared and can be identified with the string tension $\sigma$.
The other coefficients must be determined from the underlying
microscopic theory.
Examples with calculable coefficients include the $D\!=\!3\ Z(2)$ defect line
and the $D\!=\!4$ Nielsen-Olesen vortex
in a semi-classical expansion of field theory. Similar calculations in QCD
would require a quantitative understanding of quark confinement.

Asymptotic string gaps and 
the observed fine structure of the spectrum provide the most direct
tests of effective QCD string formation for large $R$.
Low frequency excitations are dominated by the first term which describes
the asymptotic spectrum of massless modes with exact $\pi/R$ string gaps.
Higher dimensional operators are expected to generate 
perturbative fine structure 
corrections in higher powers of $R^{-1}$. Matching the
coefficients of Eq.~(1) to the spectrum is a realistic goal for 
future high precision
simulations. Since it is reasonable to expect that the first few
coefficients arise from the geometric properties of the effective string, 
it is useful to introduce the Nambu-Goto (NG) action  
$\sigma\sqrt{ {\rm 1}+\partial_\mu\bm{\xi}\cdot
\partial_\mu\bm{\xi} }$ as the simplest geometric model.
The NG spectrum
with fixed end boundary conditions in $D$ dimensions was first 
calculated in Ref.~\cite{arvis} with the result
$ E_N = \sigma R ( 1 - \frac{D-2}{12\sigma R^2}\pi+ \
  \frac{2\pi N}{\sigma R^2} )^{\frac{1}{2}}.$
Although there exists a quantization problem in angular momentum
rotations around the $q\bar q$ axis at finite
$R$ values unless $D=26$~\cite{arvis}, the problem disappears 
in the large $R$ limit.
It has been widely expected that predictions at large $R$ 
from the NG model will be
similar to effective bosonic string theory with a good description
of the observed
spectrum. Deviations from the NG string at large $R$ in 
Fig.~\ref{fig:fig1} may indicate
rigidity and other missing physical properties 
of the flux line in the model.

In this work, striking confirmation of string-like flux formation
of the gluon field for large $q\bar q$ separations was presented. 
Our simulations revealed a first glimpse of a fine
structure in the large-distance spectrum, offering the
tantalizing possibility of 
understanding the nature of the QCD string in future high precision
simulations.

\begin{acknowledgments}
Throughout this extended project,
we benefited from discussions with M.~Peardon, G.~Bali, R.~Brower, 
K.~Intriligator, P.~Hasenfratz, M.~L\"uscher, C.~Michael, S.~Necco,
F.~Niedermayer, R.~Sommer, J.~Soto, P.~Weisz,
and many others.
This work was supported by the DOE, Grant No. DE-FG03-97ER40546, the 
NSF under Award PHY-0099450 and the European
Community's Human Potential Programme under contract HPRN-CT-2000-00145,
Hadrons/Lattice QCD.
\end{acknowledgments}

\end{document}